\documentclass[12p]{iopart}

\usepackage{color}
\usepackage{graphicx}
\usepackage{iopams}
\usepackage{braket}
\usepackage[utf8]{inputenc}
\usepackage{url}
\usepackage{bbold}

\begin{document}

\title[Theory of quantum FC and type-II PDC in the high-gain regime]{Theory of quantum frequency conversion and type-II parametric down-conversion in the high-gain regime}

\author{Andreas Christ$^{1}$, Benjamin Brecht$^{1}$, Wolfgang Mauerer$^{2,3}$, and Christine Silberhorn$^{1,2}$}
\address{\(^1\)Applied Physics, University of Paderborn, Warburger Straße 100, 33098 Paderborn, Germany}
\address{\(^2\)Max Planck Institute for the Science of Light,\\ G\"unther-Scharowsky Straße 1/Building 24, 91058 Erlangen, Germany}
\address{\(^3\)Siemens AG, Corporate Research and Technologies,\\ Wladimirstrasse 3, 91058 Erlangen, Germany}

\ead{andreas.christ@uni-paderborn.de}

\date{\today}

\begin{abstract}
Frequency conversion (FC) and type-II parametric down-conversion (PDC) processes serve as basic building blocks for the implementation of quantum optical experiments: type-II PDC enables the efficient creation of quantum states such as photon-number states and Einstein-Podolsky-Rosen-states (EPR-states). FC gives rise to technologies enabling efficient atom-photon coupling, ultrafast pulse gates and enhanced detection schemes. However, despite their widespread deployment, their theoretical treatment remains challenging. Especially the multi-photon components in the high-gain regime as well as the explicit time-dependence of the involved Hamiltonians hamper an efficient theoretical description of these nonlinear optical processes.

In this paper, we investigate these effects and put forward two models that enable a full description of FC and type-II PDC in the high-gain regime. We present a rigorous numerical model relying on the solution of coupled integro-differential equations that covers the complete dynamics of the process. As an alternative, we develop a simplified model that, at the expense of neglecting time-ordering effects, enables an analytical solution.

While the simplified model approximates the correct solution with high fidelity in a broad parameter range, sufficient for many experimental situations, such as FC with low efficiency, entangled photon-pair generation and the heralding of single photons from type-II PDC, our investigations reveal that the rigorous model predicts a decreased performance for FC processes in quantum pulse gate applications and an enhanced EPR-state generation rate during type-II PDC, when EPR squeezing values above 12\,dB are considered.
\end{abstract}


\maketitle

\section{Introduction}\label{sec:introduction}
A fundamental building block of quantum information and quantum communication applications are nonlinear optical processes. In experimental implementations of photonic quantum systems type-II parametric down-conversion (PDC) and frequency conversion (FC) are omnipresent. Type-II PDC enables the generation of various quantum states ranging from single photons \cite{hong_measurement_1987, uren_efficient_2004, mosley_heralded_2008, pittman_heralding_2005, migdall_tailoring_2002} over entangled photon-pairs \cite{kwiat_new_1995, kwiat_ultrabright_1999, kurtsiefer_high-efficiency_2001} up to EPR-states \cite{barnett_methods_2003, christ_probing_2011}. FC is applied for frequency translations between different wavelengths \cite{gu_photon_2012, mcguinness_quantum_2010, mejling_quantum_2012}, which enables interfaces between quantum systems, in particular atom-photon coupling \cite{tanzilli_photonic_2005, kielpinski_quantum_2011}, quantum pulse gates \cite{brecht_quantum_2011, eckstein_quantum_2011, mcguinness_theory_2011}, and efficient quantum state detection \cite{kamada_efficient_2008, pelc_long-wavelength-pumped_2011, pelc_cascaded_2012, clark_double-stage_2011, ramelow_polarization-entanglement-conserving_2012}. 

Their deployment in quantum enhanced applications requires a detailed theoretical understanding of the corresponding nonlinear interactions. A variety of models have been developed for PDC  \cite{grice_spectral_1997,wasilewski_pulsed_2006, lvovsky_decomposing_2007, brambilla_simultaneous_2004, christ_probing_2011, branczyk_non-classical_2010} and FC \cite{brecht_quantum_2011, eckstein_quantum_2011, mcguinness_theory_2011, mejling_quantum_2012,porat_two_2012}. They vary from straightforward perturbation approaches to much more rigorous treatments. The crucial issue in these derivations is firstly the fact that multi-photon effects have to be considered during the interaction, and secondly the problem that the involved electric field operators and consequently Hamiltonians do \textit{not} commute in time. In this paper we address these issues and build two theoretical models for FC and type-II PDC: a rigorous numerical model extending the theoretical framework of Kolobov \cite{kolobov_spatial_1999}, and a simplified analytical approach. Both models take into account higher-order photon number effects and are hence suitable to describe FC and type-II PDC in the high-gain regime. We analyse their performance and the quality of their predictions over a broad parameter range, which enables us to suggest in which experimental configurations a simple analytic modelling of the processes is sufficient and when the rigorous approach has to be applied.

The paper is structured into two main parts. In sections \ref{sec:fc_overview} to \ref{fc_implications} we study FC. Our investigation of this process is divided into eight sub chapters: after a short description of the basic principles of FC in section \ref{sec:fc_overview}, section \ref{sec:fc_hamiltonian} discusses the Hamiltonian of the process. The general properties of the conversion are outlined in section \ref{sec:fc_general_properties}. In section \ref{sec:fc_analytic_model_excluding_time-ordering_effects} we derive the simplified analytic solution excluding time-ordering effects. In section \ref{sec:fc_rigorous_theory_including_time-ordering_effects} we put forward the rigorous approach relying on the solution of coupled integro-differential equations. The differences between the two models are quantified in section \ref{sec:fc_comparison_between_simplified_analytical_and_rigorous_approach}. Finally, in section \ref{fc_implications}, we elaborate on the impacts of our work on the design and performance of FC processes for quantum enhanced applications. The same reasoning is then applied to the process of type-II PDC in sections \ref{sec:pdc_overview} onward. Section \ref{sec:conclusion} concludes the paper and summarizes our findings. \ref{app:FC_canonical_transformation_conditions} to \ref{app:pdc_numerical_implementation} contain additional information and further calculations.

\section{Frequency conversion (FC): overview}\label{sec:fc_overview}
\begin{figure}[htp]
\begin{center}
    \includegraphics[width=0.9\textwidth]{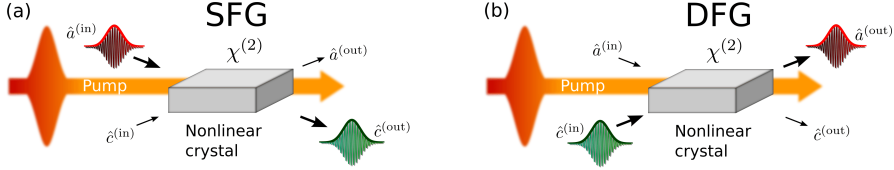}
\end{center}
\caption{Sketch of the FC process. Mediated by the strong pump field and the \(\chi^{(2)}\)-nonlinearity of the medium parts of the fields either an input field \(\hat{a}^{\mathrm{(in)}}\) is converted via SFG to \(\hat{c}^{\mathrm{(out)}}\) (a) or via the process of DFG a field in \(\hat{c}^{\mathrm{(in)}}\) is converted to \(\hat{a}^{\mathrm{(out)}}\).}

\label{fig:frequency_conversion_setup}
\end{figure}

A general FC process is sketched in figure \ref{fig:frequency_conversion_setup}. Mediated by the nonlinearity of the crystal and a strong pump beam two input fields \(\hat{a}^{\mathrm{(in)}}\) and \(\hat{c}^{\mathrm{(in)}}\) are interconverted into two output fields \(\hat{a}^{\mathrm{(out)}}\) and \(\hat{c}^{\mathrm{(out)}}\). This FC process is more commonly known as sum frequency generation (SFG), when the input beam in combination with the pump beam generates an output field at a higher frequency \(\omega_{out} = \omega_{in} + \omega_{p}\) (figure \ref{fig:frequency_conversion_setup}(a)), or difference frequency generation (DFG), when a field with frequency \(\omega_{out} = \omega_{in} - \omega_{p}\) is created (figure \ref{fig:frequency_conversion_setup}(b)).\footnote{In classical nonlinear optics, DFG is understood as a stimulated process. The bright pump field has the highest frequency and the process is seeded with a weak input field, which is enhanced through continuous conversion of pump photons. We, in contrast, assume a weak input field, which has the highest frequency and the ‘seed’ is the bright pump field (see \cite{brecht_quantum_2011} for details).}

The distinction between SFG and DFG arises via the input wave that is fed in either the \(\hat{a}^{\mathrm{(in)}}\) or \(\hat{c}^{\mathrm{(in)}}\) port. In the scope of this paper \(\hat{a}^{\mathrm{(in)}} \rightarrow \hat{c}^{\mathrm{(out)}}\) depicts a SFG process and \(\hat{c}^{\mathrm{(in)}} \rightarrow \hat{a}^{\mathrm{(out)}}\) labels DFG. Each crystal configuration always supports both processes simultaneously and we refer to the overall system as frequency conversion.

In this paper, we go beyond the standard monochromatic single-mode description of FC and consider a multitude of frequencies interacting with each other during the FC process. This becomes especially important for ultrafast pulsed FC experiments where ultrafast light pulses with spectral bandwidth of several nanometers interact with each other \cite{ porat_two_2012, kuzucu_time-resolved_2008, brecht_quantum_2011, eckstein_quantum_2011}. In the following chapters, we derive the properties of this transformation and compare the accuracy of different theoretical models.

\section{FC: Hamiltonian}\label{sec:fc_hamiltonian}
We first define the electric field operators of an optical wave inside a nonlinear medium as \cite{blow_continuum_1990}
\begin{eqnarray}
    \nonumber
    \fl \,\,\,\,\,\, \hat{E}^{\left(+\right)}(z,t) = \hat{E}^{\left(-\right)\dagger}(z,t)
    \\ =  \imath \left(\frac{\hbar \, c}{4 \pi \epsilon_0  A \, n^3(k_{0}) }\right)^{\frac{1}{2}} \int \mathrm dk \,  \sqrt{\omega(k)} \exp\left[\imath(k z - \omega(k) t)\right]\hat{a}(k),
    \label{eq:electric_field_1D}
\end{eqnarray}
where \(A\) labels the transverse quantization area in the material \cite{loudon_quantum_2000}. We use the slowly varying envelope approximation, i.e. the bandwidth \(\Delta \omega\) of the considered electric fields is small compared to their central frequency \(\omega_0\) (\(\Delta \omega \ll \omega_0\)) and hence treat the dispersion term in front of the integral \(n(k_{0})\) in \eref{eq:electric_field_1D} as a constant, using the value at the central wave vector \(k_0\). This approximation is justified since, in the remainder of this paper, we only consider electric fields not too broad in frequency, compared to their central frequency, and take into account the rather flat dispersion in nonlinear crystals. Finally \(\hat{a}(k)\) is the standard single-photon annihilation operator that destroys a single-photon in mode \(k\) and obeys
\begin{equation}
\left[\hat{a}(k),\hat{a}({k')}^\dagger\right] = \delta(k-k') \qquad
\left[\hat{a}(k),\hat{a}({k')}\right] = 0.
\end{equation}
In this paper, we restrict ourselves to electric fields in one dimension. This means we assume a fully collinear propagation of the interacting fields along one axis in a single spatial mode, as e.g. given inside a waveguiding structure, since a three dimensional treatment does not offer further physical insight into the properties of the process and complicates the calculations.

The Hamiltonian of the FC process consists of two parts: The Hamiltonian  \(H_0^{(x)}\) describes the free propagation of the electromagnetic waves through the medium for each of the involved fields \cite{kolobov_spatial_1999},
\begin{equation}
    \hat{H}_0^{(x)}(t) =  \frac{2 \epsilon_0 A \, n_{k_0^{x}}^{3}}{c} \int \mathrm d z \, \hat{E}_x^{(-)}(z,t) \hat{E}_x^{(+)}(z,t),
\label{eq:fc_free_space_hamiltonian}
\end{equation}
where \(x\) represents either the pump \(p\) or the two interconverted fields \(a, c\) and the factor in front of the Hamiltonian appears due to the normalization of the electric fields operators in \eref{eq:electric_field_1D} \cite{kolobov_spatial_1999}. The interaction Hamiltonian of the frequency conversion process is given by \cite{kumar_quantum_1990, odonnell_time-resolved_2009, brecht_quantum_2011, eckstein_quantum_2011}
\begin{equation}
       \fl \qquad \qquad  \hat{H}_I^{(FC)}(t) = \epsilon_0 \,  \int \mathrm dz \, \chi^{(2)}(z) \, E_p^{(+)}(z,t)  \hat{E}_a^{(+)}(z,t) \hat{E}_c^{(-)}(z,t) + h.c. \,.
    \label{eq:fc_interaction_hamiltonian}
\end{equation}
\(E_p^{(+)}(z,t)\) labels the pump field driving the FC process and \(\hat{E}_a^{(+)}(z,t), \hat{E}_c^{(-)}(z,t)\) are the two fields that are interconverted. In the derivation of this Hamiltonian, we used the rotating wave approximation and hence only consider the FC terms of the nonlinear optical process while neglecting the PDC and further nonlinear interactions. We further assume that the nonlinearity is constant throughout the material. The pump field driving the frequency conversion process is a strong optical wave and is consequently treated as a classical wave:
\begin{equation}
    \fl \qquad \qquad  E_p^{\left(+\right)}(z,t) = E_p^{\left(-\right)*}(z,t)=
    \mathcal{A}_p \int \mathrm dk \, \alpha\left[\omega(k)\right] \exp\left[\imath(k z - \omega(k) t)\right].
    \label{eq:classical_pump_field}
\end{equation}
Here \(\mathcal{A}_p\) labels the pump amplitude and \(\alpha\left[\omega(k)\right]\) its spectral distribution ranging from \(\delta(\omega - \omega_c)\) for cw-laser sources up to more complicated forms in the case of pulsed laser systems. We further assume that the pump field is not depleted during propagation through the crystal since only a minor part of the strong pump beam is lost during the FC process. 

Combining \eref{eq:fc_free_space_hamiltonian} and \eref{eq:fc_interaction_hamiltonian}, the FC process is described by the overall Hamiltonian:
\begin{equation}
    \hat{H}_{FC}(t) = \hat{H}_0^{(a)}(t) + \hat{H}_0^{(c)}(t) + \hat{H}_I^{(FC)}(t).
    \label{eq:fc_hamiltonian}
\end{equation}
There are a variety of different constants involved in the definition of the FC Hamiltonian in \eref{eq:fc_hamiltonian} (see \eref{eq:electric_field_1D}, \eref{eq:fc_free_space_hamiltonian}, \eref{eq:classical_pump_field} and \eref{eq:fc_interaction_hamiltonian}). However, these do not change the qualitative behaviour of the process. In the remainder of this paper, we merge all of them into a coupling value depicting the overall efficiency of the frequency conversion process rendering the presented calculations independent of individual notations.

\section{FC: general properties}\label{sec:fc_general_properties}
From the FC Hamiltonian in \eref{eq:fc_hamiltonian}, we are able to calculate the unitary transformation \cite{sakurai_modern_1993, nielsen_quantum_2004, barnett_methods_2003} generated by the FC process, with the help of the Schr\"odinger equation \cite{sakurai_modern_1993, nielsen_quantum_2004}:
\begin{equation}
    \hat{U}_{FC} = \mathcal{T} \exp\left[-\frac{\imath}{\hbar} \int \mathrm dt \, \hat{H}_{FC}(t) \right].
    \label{eq:fc_unitary_transformation}
\end{equation}
This unitary \(\hat{U}_{FC}\) describes the transformation of the beams during propagation through the crystal, i.e. the transformation from the light fields impinging on the crystal \(\hat{a}^{\mathrm{(in)}}\) and \(\hat{c}^{\mathrm{(in)}}\) to the output fields \(\hat{a}^{\mathrm{(out)}}\) and \(\hat{c}^{\mathrm{(out)}}\) (see figure \ref{fig:frequency_conversion_setup}).

In \eref{eq:fc_unitary_transformation} the \textit{time-ordering} operator \(\mathcal{T}\) is crucial, because the electric field operators in \(\hat{H}_{FC}(t)\) are time-dependent. In turn, the Hamiltonian does not commute at different points in time, which renders finding a solution difficult. Nevertheless the structure of \eref{eq:fc_unitary_transformation} gives valuable insights into the properties of the system, because the Hamiltonian in \eref{eq:fc_unitary_transformation} is bilinear in its electric field operators if the pump is treated as a classical wave. The solution hence takes, in the Heisenberg picture, the form of a linear unitary Bogoliubov transformation \cite{braunstein_quantum_2005, campos_quantum-mechanical_1989, ekert_relationship_1991, braunstein_squeezing_2005}:\footnote{Solving the process via the Heisenberg equation of motion (see section \ref{sec:fc_rigorous_theory_including_time-ordering_effects}) yields a linear operator-valued differential equation. Linear operator-valued differential equations are solved by a linear Bogoliubov transformation \cite{braunstein_quantum_2005, braunstein_squeezing_2005}, i.e. the solution will be of the form \eref{eq:fc_general_solution}. Furthermore with the help of the Bloch-Messiah reduction, discussed in \ref{app:FC_canonical_transformation_conditions}, we are able to rewrite the general linear Bogoliubov transformation  into the form depicted in \eref{eq:fc_general_broadband_solution}.} 
\begin{eqnarray}
    \nonumber
    &\hat{a}^{\mathrm{(out)}}(\omega) = \int \mathrm d \omega' U_a(\omega, \omega')\, \hat{a}^{\mathrm{(in)}}(\omega') + \int \mathrm d \omega' V_a(\omega, \omega')\, \hat{c}^{\mathrm{(in)}}(\omega'), \\
    &\hat{c}^{\mathrm{(out)}}(\omega) = \int \mathrm d \omega' U_c(\omega, \omega')\, \hat{c}^{\mathrm{(in)}}(\omega') - \int \mathrm d \omega' V_c(\omega, \omega')\, \hat{a}^{\mathrm{(in)}}(\omega').
    \label{eq:fc_general_solution}
\end{eqnarray}
In the scope of this paper, the conversion from \(\hat{a}^{\mathrm{(in)}} \rightarrow \hat{c}^{\mathrm{(out)}}\) depicts SFG and \(\hat{c}^{\mathrm{(in)}} \rightarrow \hat{a}^{\mathrm{(out)}}\) DFG. The functions \(U_{a/c}(\omega, \omega')\) in  \eref{eq:fc_general_solution} define which parts of the different frequencies of the input beams pass the crystal unperturbed, whereas the  \(V_{a/c}(\omega, \omega')\) functions give the portions of the beams which are converted.

In order to unravel the underlying structure, we use the constraint that the FC process is a unitary transformation and hence \eref{eq:fc_general_solution} is a canonical operator transformation \cite{braunstein_quantum_2005, ekert_relationship_1991, braunstein_squeezing_2005}. This imposes several conditions on the properties of the solution, which we study in detail in \ref{app:FC_canonical_transformation_conditions}. Under this constraint and with the help of the Bloch-Messiah decomposition, we rewrite \eref{eq:fc_general_solution} as
\begin{eqnarray}
    \nonumber
    &\hat{A}_k^{\mathrm{(out)}} = \cos(r_k)\, \hat{A}_k^{\mathrm{(in)}} + \sin(r_k)\, \hat{C}_k^{\mathrm{(in)}},\\
    &\hat{C}_k^{\mathrm{(out)}} = \cos(r_k)\, \hat{C}_k^{\mathrm{(in)}} - \sin(r_k)\, \hat{A}_k^{\mathrm{(in)}},
    \label{eq:fc_general_broadband_solution}
\end{eqnarray}
where \(\hat{A}_k\) and \(\hat{C}_k\) are broadband single-photon destruction operators \cite{rohde_spectral_2007} defined as
\begin{eqnarray}
    \nonumber
    \fl \qquad &\hat{A}_k^{\mathrm{(out)}} = \int \mathrm d \omega \, \varphi_k(\omega)\, \hat{a}^{\mathrm{(out)}}(\omega), \qquad
    \hat{C}_k^{\mathrm{(out)}} = \int \mathrm d \omega \, \xi_k(\omega) \, \hat{c}^{\mathrm{(out)}}(\omega), \\
    \fl \qquad  &\hat{A}_k^{\mathrm{(in)}} = \int \mathrm d \omega \, \psi_k(\omega) \,\hat{a}^{\mathrm{(in)}}(\omega), \qquad \,\,\,\,\,
    \hat{C}_k^{\mathrm{(in)}} = \int \mathrm d \omega \, \phi_k(\omega) \, \hat{c}^{\mathrm{(in)}}(\omega).
    \label{eq:broadband_mode_operators}
\end{eqnarray}
In essence, an ultrafast FC process converts ultrafast optical pulses given by the mode shapes \(\psi_k(\omega)\) and \(\phi_k(\omega)\) into the pulse modes \(\varphi_k(\omega)\) and \(\xi_k(\omega)\). The conversion efficiency of each mode \(k\) is given via \(\sin^2(r_k)\).

Note that this principle also provides the underlying concept for considering the overall FC process as a quantum pulse gate or quantum pulse shaper \cite{eckstein_quantum_2011, brecht_quantum_2011, mcguinness_theory_2011}. Depending on the efficiency of the process it transmits the incoming pulses unperturbed or switches them via FC. The crucial parameters of this transformation are firstly the conversion efficiencies  \(\sin^2(r_k)\) and secondly the ultrafast mode shapes \(\varphi_k(\omega)\), \(\xi_k(\omega)\), \(\psi_k(\omega)\) and \(\phi_k(\omega)\) that define the range of frequencies that are interconverted.

\section{FC: analytic model excluding time-ordering effects}\label{sec:fc_analytic_model_excluding_time-ordering_effects}
Unfortunately, it is not trivial to evaluate \eref{eq:fc_unitary_transformation} due to the time-ordering operator \(\mathcal{T}\) in front of the exponential function. Neglecting these time-ordering effects, however, enables us to build an analytical model of FC, which is highly beneficial, for practical purposes, since it enables quick and straightforward access to the process properties:\footnote{For a discussion of the physical meaning of the time-ordering operator in the context of PDC and FC please have a look at the works of Bra\'{n}czyk \cite{branczyk_non-classical_2010, branczyk_time_2011} and \cite{christ_theory_2013}.}
\begin{equation}
    \hat{U}_{FC} = \exp\left[-\frac{\imath}{\hbar} \int \mathrm dt \, \hat{H}_I^{(FC)}(t) \right].
    \label{eq:fc_no_time_ord_unitary_transformation}
\end{equation}
This formula is identical to \eref{eq:fc_unitary_transformation} except that we drop the time ordering operator \(\mathcal{T}\) and work in the interaction picture. It enables us to directly perform the time integration in the exponential function. While it is possible to perform this calculation using the electric field operators as defined in \eref{eq:electric_field_1D} we are able to considerably simplify these calculations by working with the electric field operators in the \(\omega\)-representation \cite{blow_continuum_1990}:
\begin{eqnarray}
    \nonumber
    \fl \qquad \hat{E}^{\left(+\right)}(z,t) = \hat{E}^{\left(-\right)\dagger}(z,t) \\
    = \imath \left(\frac{\hbar}{4 \pi \epsilon_0  c\, A \, n(\omega_{0})}\right)^{\frac{1}{2}} \int \mathrm d\omega \, \sqrt{\omega} \exp\left[\imath(k(\omega) z - \omega t)\right]\hat{a}(\omega).
    \label{eq:fc_electric_field_1D_frequency}
\end{eqnarray}
We also perform our calculations in the interaction picture. This means we move into a new reference frame where the effects of free propagation are not present and hence do not need to consider the free propagation Hamiltonians. Finally, we assume a crystal featuring a constant \(\chi^{(2)}\)-nonlinearity extending from \(-\frac{L}{2}\) to \(\frac{L}{2}\). After a straightforward calculation we obtain
\begin{eqnarray}
    \fl \qquad \hat{U}_{FC} = \exp\left[-\frac{\imath}{\hbar}\left(\int \mathrm d \omega_a \, \int \mathrm d \omega_c \, f(\omega_a, \omega_c) \hat{a}(\omega_a) \hat{c}^\dagger(\omega_c) + h.c. \right) \right],
    \label{eq:fc_no_time_ordering}
\end{eqnarray}
where \(f(\omega_a, \omega_c)\) is defined as:
\begin{eqnarray}
 f(\omega_a, \omega_c) =   B \, \alpha\left(\omega_c - \omega_a\right) \, \mathrm{sinc}\left(\frac{\Delta k(\omega_a, \omega_c) L}{2}\right).
\label{eq:fc_joint_spectral_amplitude}
\end{eqnarray}
Here we merged all constants into the overall factor \(B\) and \(\Delta k(\omega_a, \omega_c) = k_p(\omega_c - \omega_a) + k_a(\omega_a) - k_c(\omega_c)\). Details of this calculation are given in \cite{brecht_quantum_2011}.

With the help of the singular-value-decomposition (SVD) theorem \cite{law_continuous_2000}, we recast this solution in the broadband mode formalism presented in \eref{eq:fc_general_broadband_solution}. At first we diagonalize the Hamiltonian by decomposing the exponent in \eref{eq:fc_no_time_ordering}, via a Schmidt decomposition, as
\begin{eqnarray}
    \nonumber
    -\frac{\imath}{\hbar} f(\omega_a, \omega_c) &= \sum_k (-r_k) \psi_k(\omega_a) \phi_k^*(\omega_c), \\
    -\frac{\imath}{\hbar} f^*(\omega_a, \omega_c) &= \sum_k r_k \psi_k^*(\omega_a) \phi_k(\omega_c).
    \label{eq:fc_schmidt_decomposition}
\end{eqnarray}
Here both \(\{\psi_k(\omega_a)\}\) and \(\{\phi_k(\omega_c)\}\) each form a complete set of orthonormal functions and  \(r_k \in \mathcal{R}^+\). Employing equation \eref{eq:fc_schmidt_decomposition} we rewrite the unitary FC transformation in \eref{eq:fc_no_time_ordering} as
\begin{eqnarray}
    \nonumber
    \fl \qquad \hat{U}_{FC} = \exp\left[\sum_k (-r_k) \int \mathrm d \omega_a \, \psi_k(\omega_a) \hat{a}(\omega_a) \int \mathrm d \omega_c \phi_k^*(\omega_c) \hat{c}^\dagger(\omega_c) \right. \\
    \left . + r_k \int \mathrm d \omega_a \psi_k^*(\omega_a) \hat{a}^\dagger(\omega_a) \int \mathrm d \omega_c \phi_k(\omega_c) \hat{c}(\omega_c)\right].
\end{eqnarray}
With the help of the broadband mode operators defined in \eref{eq:broadband_mode_operators}, it takes on the form 
\begin{eqnarray}
    \nonumber
    \hat{U}_{FC} &= \exp\left[\sum_k (-r_k)  \left(\hat{A}_k \hat{C}^\dagger_k - \hat{A}_k^\dagger \hat{C}_k \right)\right]\\
    &= \bigotimes_k \exp\left[ (-r_k) \left(\hat{A}_k \hat{C}^\dagger_k - \hat{A}_k^\dagger \hat{C}_k \right) \right].
    \label{eq:fc_no_time_ordering_broadband}
\end{eqnarray}
In the Heisenberg pictures, it reads \cite{campos_quantum-mechanical_1989}
\begin{eqnarray}
    \nonumber
    &\hat{A}_k^{\mathrm{(out)}} = \cos(r_k) \hat{A}_k^{\mathrm{(in)}} + \sin(r_k) \hat{C}_k^{\mathrm{(in)}},\\
    &\hat{C}_k^{\mathrm{(out)}} = \cos(r_k) \hat{C}_k^{\mathrm{(in)}} - \sin(r_k) \hat{A}_k^{\mathrm{(in)}}.
    \label{eq:fc_no_time_ordering_broadband_input_output_relation}
\end{eqnarray}
This simplified analytic model features exactly the structure required by the canonical commutation relations discussed in section \ref{sec:fc_general_properties}. Only the additional fact that the input modes and output modes in this simplified model are always of identical shape differs from the general solution \eref{eq:fc_general_broadband_solution}.

It is evident that this treatment ignoring time-ordering effects enables a straightforward analytic solution of the FC process. In some cases, even the SVD can be performed analytically and hence no computational effort is required at all \cite{uren_photon_2003}. This enables the efficient engineering and design of frequency conversion processes as long as the applied approximations hold.

\section{FC: rigorous theory including time-ordering effects}\label{sec:fc_rigorous_theory_including_time-ordering_effects}
In order to obtain a rigorous solution of FC, we cannot neglect the effects of time-ordering. For this reason we change our analysis method and regard the FC process in the Heisenberg picture. This approach has already been utilized for FC in nonlinear optical fibers in \cite{mcguinness_quantum_2010, mcguinness_theory_2011} and is common for PDC \cite{kolobov_spatial_1999, brambilla_simultaneous_2004, wasilewski_pulsed_2006, lvovsky_decomposing_2007, mauerer_colours_2008, caspani_tailoring_2010,   wasilewski_pairwise_2006, dayan_theory_2007, gatti_multiphoton_2003}. In order to solve the corresponding Heisenberg equations of motion, we adapt the work of Kolobov on type-I PDC in \cite{kolobov_spatial_1999} to FC, which provides a rigorous and complete solution of the FC process. To simplify the calculations we redefine the electric field operators in \eref{eq:electric_field_1D} according to \cite{kolobov_spatial_1999} by dropping the constants, which would otherwise complicate the formulae without adding new insights:
\begin{eqnarray}
    \nonumber
    &\hat{a}(z,t) = \frac{1}{\sqrt{2 \pi k_0}} \int \mathrm dk\,\sqrt{\omega(k)} \exp\left[\imath \left(k z - \omega(k) t \right) \right] \hat{a}(k)
    \label{eq:fc_normalized_electric_field_a},\\
    &\hat{c}(z,t) = \frac{1}{\sqrt{2 \pi k_0}} \int \mathrm dk\,\sqrt{\omega(k)} \exp\left[\imath \left(k z - \omega(k) t \right) \right] \hat{c}(k)
    \label{eq:fc_normalized_electric_field_c}.
\end{eqnarray}
The Heisenberg equation of motion for \(\hat{a}(z,t)\) reads
\begin{equation}
    \frac{d}{dt} \hat{a}(z,t) = \frac{\imath}{\hbar} \left[\hat{H}_{FC}(t), \hat{a}(z,t)\right].
    \label{eq:fc_temporalEvolution_2}
\end{equation}
Adapting the work of \cite{kolobov_spatial_1999} we are obtain two operator-valued integro-differential equations describing the FC process including time-ordering effects:
\begin{eqnarray}
    \nonumber
    \fl \qquad \frac{\partial}{\partial z}\hat{\epsilon}_a(z,\omega) =  -\frac{\imath}{\hbar} D^*
    \int \mathrm d\omega' \epsilon_p^{(-)}(z,\omega' - \omega)  \exp\left[-\imath \Delta k(\omega, \omega') z \right] \hat{\epsilon}_c(z,\omega'),\\
    \fl \qquad \frac{\partial}{\partial z} \hat{\epsilon}_c(z,\omega) =  
    -\frac{\imath}{\hbar} D \int \mathrm d\omega' \epsilon_p^{(+)}(z,\omega - \omega') 
     \exp\left[\imath \Delta k(\omega', \omega) z \right] \hat{\epsilon}_a(z,\omega').
\label{eq:fc_diff_eq_final_1}
\end{eqnarray}
Here we separated the interaction from the propagation effects by transforming our operators into the interaction picture, similar to the analytic solution in section \ref{sec:fc_analytic_model_excluding_time-ordering_effects}. For this purpose we introduced the electric fields
\begin{eqnarray}
    \nonumber
    \hat{\epsilon}_a(z, \omega) = \hat{a}(z,\omega) \exp\left[-\imath k_a(\omega) z\right],\\
    \nonumber
    \hat{\epsilon}_c(z, \omega) = \hat{c}(z,\omega) \exp\left[-\imath k_c(\omega) z\right],\\
    \epsilon_p^{(-)}(z, \omega) = E_p^{(-)}(z,\omega) \exp\left[\imath k_p(\omega) z\right].
    \label{eq:group_velocity_transformation}
\end{eqnarray}
Finally we used the abbreviation \(\Delta k(\omega, \omega') = k_p(\omega' - \omega) - k_c(\omega') + k_a(\omega) \). Note that, between the two formulae in \eref{eq:fc_diff_eq_final_1}, the variables \(\omega\) and \(\omega'\) in the \(\Delta k\) and \(\epsilon_p\) functions are flipped. By defining
\begin{equation}
    f(\omega, \omega', z) =  - \frac{\imath}{\hbar} D^* \epsilon_p^{(-)}(z, \omega' - \omega)\exp\left[-\imath \Delta k(\omega, \omega')z\right],
\end{equation}
we may write \eref{eq:fc_diff_eq_final_1} in a more compact notation:
\begin{eqnarray} \nonumber
    \frac{\partial}{\partial z}\hat{\epsilon}_a(z,\omega) =  
    \int \mathrm d\omega' f(\omega, \omega', z) \,  \hat{\epsilon}_c(z,\omega'), \\
    \frac{\partial}{\partial z} \hat{\epsilon}_c(z,\omega) =  
     - \int \mathrm d\omega' f^*(\omega', \omega, z) \, \hat{\epsilon}_a(z,\omega').
\label{eq:fc_diff_eq_final_2}
\end{eqnarray}
A detailed derivation of \eref{eq:fc_diff_eq_final_2} is given in \cite{christ_theory_2013}.

\subsection{Solving the differential equations}\label{sec:fc_solving_the_diff_eq}
In order to obtain the dynamics of the FC process, the differential equations in  \eref{eq:fc_diff_eq_final_2} have to be solved. Usually operator-valued differential equations cannot readily be evaluated and, in the case of FC, this is complicated by the fact that we have to solve integro-differential equations, since we consider the conversion of many frequencies simultaneously. However, note that \eref{eq:fc_diff_eq_final_2} is linear in its operators and hence classical solution methods like the split-step Fourier inversion method have been applied \cite{mcguinness_theory_2011, wasilewski_pulsed_2006, lvovsky_decomposing_2007}. In the special case of a cw-pump laser, the integral in \eref{eq:fc_diff_eq_final_2} vanishes and it is even possible to find analytic solutions \cite{kolobov_spatial_1999}.

In this paper we apply a different approach --- introduced by Mauerer in \cite{mauerer_colours_2008} --- exploiting the fact that the structure of the solution is already known: it is a linear operator transformation \eref{eq:fc_general_solution}. Using \eref{eq:fc_general_solution} and \eref{eq:fc_diff_eq_final_2} we obtain two pairs of classical integro-differential equations \cite{brambilla_simultaneous_2004}:
\begin{eqnarray}
    \nonumber
    \frac{\partial}{\partial z} U_a(z,\omega, \omega'') = - \int \mathrm d \omega' f(\omega, \omega', z) V_c(z, \omega', \omega''), \\
    \frac{\partial}{\partial z} V_c(z,\omega, \omega'') = \int \mathrm d \omega' f^*(\omega', \omega, z) U_a(z, \omega', \omega'') 
    \label{eq:classical_differential_equations_1}
\end{eqnarray}
and
\begin{eqnarray}
    \nonumber
    \frac{\partial}{\partial z} U_c(z,\omega, \omega'') = - \int \mathrm d \omega' f^*(\omega', \omega, z) V_a(z, \omega', \omega''), \\
    \frac{\partial}{\partial z} V_a(z,\omega, \omega'') = \int \mathrm d \omega' f(\omega, \omega', z) U_c(z, \omega', \omega'')
    \label{eq:classical_differential_equations_2}
\end{eqnarray}
which cover the complete dynamics of the FC process.

We solve these coupled integro-differential equations using an iterative approach. For the pair in \eref{eq:classical_differential_equations_1} this means we first formally integrate both differential equations along \(z\), where we assume a medium of length \(L\), as in in the analytic solution discussed in section \ref{sec:fc_analytic_model_excluding_time-ordering_effects},
\begin{eqnarray}
    \nonumber
    U_a(z,\omega, \omega'') = \delta(\omega-\omega') - \int^{\frac{L}{2}}_{-\frac{L}{2}} \mathrm d z \, \int \mathrm d \omega' f(\omega, \omega', z) V_c(z, \omega', \omega''),\\
    V_c(z,\omega, \omega'') = \int^{\frac{L}{2}}_{-\frac{L}{2}} \mathrm d z \, \int \mathrm d \omega' f^*(\omega', \omega, z) U_a(z, \omega', \omega'').
    \label{eq:integrated_classical_differential_equations_1}
\end{eqnarray}
Here we also included our knowledge about the initial solution. If  no interaction takes place the process is described by the identity operation \(\hat{a}^{\mathrm{(out)}}(\omega) = \hat{a}^{\mathrm{(in)}}(\omega)\) and \(\hat{c}^{\mathrm{(out)}}(\omega) = \hat{c}^{\mathrm{(in)}}(\omega)\) from which follows
\begin{eqnarray}
    \nonumber
    U_a(z, \omega, \omega'') = U_c(z, \omega, \omega'') = \delta(\omega - \omega''),\\
    V_a(z, \omega, \omega'') = V_c(z, \omega, \omega'') = 0
    \label{eq:fc_inital_solution}
\end{eqnarray}
Starting with the initial solution for \(U_a(z, \omega, \omega'')\) we then perform the two integrations in \eref{eq:integrated_classical_differential_equations_1} and obtain a preliminary \(V_c(z, \omega, \omega'')\). This is then used to obtain a new \(U_a(z, \omega, \omega'')\). We repeat this iterative procedure till the functions converge. 

The same method is applied to the second set of differential equations defining \(U_c(z, \omega, \omega'')\) and \(V_a(z, \omega, \omega'')\), which gives us the complete time-ordered solution of the FC process. The implementation of this algorithm is discussed in \ref{app:fc_numerical_details}, where we also elaborate on the numerical accuracy of the applied method.\footnote{The program code, written in Python, can be downloaded from the publications section on our website. The current url is \url{http://physik.uni-paderborn.de/ag/ag-silberhorn/publications/}.}

\section{FC: comparison between simplified analytical and rigorous approach}\label{sec:fc_comparison_between_simplified_analytical_and_rigorous_approach}
In order to quantify the discrepancies between the two approaches presented in sections \ref{sec:fc_analytic_model_excluding_time-ordering_effects} and \ref{sec:fc_rigorous_theory_including_time-ordering_effects} we simulate an almost uncorrelated FC process where only the first optical mode \(r_k\) is strongly excited \cite{brecht_quantum_2011}, since this is the case where the differences between the different models are most prominent. Furthermore, this configuration also corresponds to quantum pulse gate operation of frequency conversion \cite{brecht_quantum_2011, eckstein_quantum_2011, mcguinness_theory_2011}. The exact simulation parameters are given in \ref{app:fc_simulated_frequency_conversion_process}.

\begin{figure}[htb]
    \begin{center}
        \includegraphics[width=\textwidth]{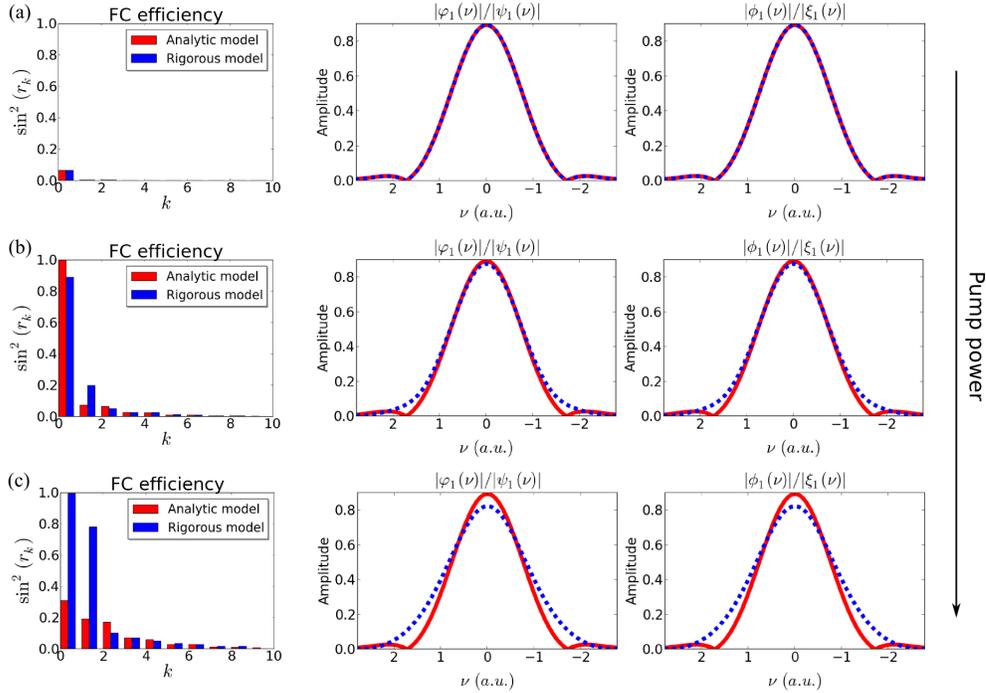}
    \end{center}
    \caption{Comparison between the rigorous and the analytical approach in uncorrelated few-mode ultrafast FC. In the low-conversion regime, presented in (a), both approaches evaluate to identical results (\(6.4 / 6.3 \%\) conversion efficiency in the first mode in the analytic / rigorous model). Approaching unit efficiency in (b) the two approaches start to show differences (\(100 / 89 \%\) conversion efficiency in the first mode), which become significant when optical gains beyond unity are considered in (c) (\(30 / 99 \%\) conversion efficiency in the first mode).}
    \label{fig:fc_time-ordering_result_uncorrelated}
\end{figure}

The obtained FC efficiencies \(\sin^2(r_k)\) and pulse shapes are displayed in figure \ref{fig:fc_time-ordering_result_uncorrelated}. The figures in the column on the left show the conversion efficiencies \(\sin^2(r_k)\), whereas the two columns on the right present the corresponding mode functions \(\varphi_1(\nu), \psi_1(\nu), \phi_1(\nu)\) and \(\xi_1(\nu)\) for the first optical mode featuring the highest conversion efficiency, where \(\varphi_1(\nu)\) and  \(\psi_1(\nu)\) as well as \(\phi_1(\nu)\) and \(\xi_1(\nu)\) are of identical shape.

In the low conversion case, i.e. in the case of low pump powers, depicted in Fig. \ref{fig:fc_time-ordering_result_uncorrelated} (a), both models yield identical results. When, with strong pump beams, unit conversion efficiency is approached, as shown in figure \ref{fig:fc_time-ordering_result_uncorrelated} (b), first discrepancies between the different models start to appear. The mode functions derived from the rigorous model show a small broadening and the conversion efficiency in the very first mode rises slower as expected from the analytical approach. The second mode however rises faster. This means the whole systems moves from a single-mode to a multi-mode operation. Significant differences between the two theories occur when we choose to use even higher pump powers as depicted in \ref{fig:fc_time-ordering_result_uncorrelated} (c). The conversion efficiencies, in the rigorous model, presented in \ref{fig:fc_time-ordering_result_uncorrelated}(c), remain at unit conversion efficiency once they reach this value. In contrast they drop in the analytical model. Furthermore, the rigorous model predicts a significant broadening of the corresponding mode shapes in the high-gain regime.

\section{FC: implications for experimental implementations}\label{fc_implications}
In summary, the analytical model accurately describes FC in the low-gain regime. In the high-gain regime, when conversion efficiencies about unity are reached, complicated non-trivial deviations from the analytical model have to be taken into account and a rigorous treatment of FC is necessary. For most experimental setups and applications it is hence perfectly justified to apply the simplified analytic model to minimize the computational effort, as long as its limitations are kept in mind.

Especially affected, however, are FC processes that serve as quantum pulse gates \cite{brecht_quantum_2011, eckstein_quantum_2011, mcguinness_theory_2011}. In theory, a perfect quantum pulse gate converts a single optical mode with unit efficiency. However, as is evident from figure \ref{fig:fc_time-ordering_result_uncorrelated}, the time-ordering effects move the FC process from the single-mode regime towards a more multi-mode behaviour. This effect fundamentally limits the gate performance at high conversion efficiencies. One could, in principle, use advanced engineering techniques such as hypergrating structures to reduce the multi-mode character in the state \cite{branczyk_engineered_2011}, yet still the time-ordering effects seem to remain a fundamental limitation. Whether or not it is actually possible to engineer single-mode quantum pulse gates including the effects of time-ordering remains an open research question.

Furthermore, our rigorous model shows that the pump power dependence of FC with only a few optical modes \(r_k\), strongly deviates from the expected sinusoidal pattern. According to the simplified model, one would expect that with increasing pump power the conversion efficiency shows a \(\sin^2\) dependence on the pump amplitude. However, according to our rigorous model strong deviations from this behaviour appear, when uncorrelated FC processes are considered. Instead of simply decreasing after unit conversion efficiency is reached, the conversion efficiency of the first optical mode remains at unity despite rising pump powers. While this quite unexpected behaviour is not present in a FC processes featuring a multitude of optical modes \(r_k\) it has to be taken into account when single- or few-mode frequency conversion experiments are performed.

\section{Parametric down-conversion (PDC): overview}\label{sec:pdc_overview}
To date, most experimental implementations of type-II PDC aim for the generation of photon-pairs \cite{hong_measurement_1987}. This is achieved by driving the type-II PDC process with very low pump powers, where the whole system can be treated using first-order perturbation theory \cite{grice_spectral_1997}. However, considering the demand for high photon-pair generation rates in current quantum optical experiments, higher and higher pump powers are applied in type-II PDC experiments \cite{eckstein_highly_2011, gerrits_generation_2011}. In this regime, perturbation approaches focusing on photon-pair generation are not sufficient any more and higher-order effects have to be taken into account. In order to mathematically describe type-II PDC in the high-gain regime we extend our theoretical framework for FC processes to type-II PDC, covering both degenerate and non-degenerate configurations.

\begin{figure}[htp]
\begin{center}
    \includegraphics[width=0.5\textwidth]{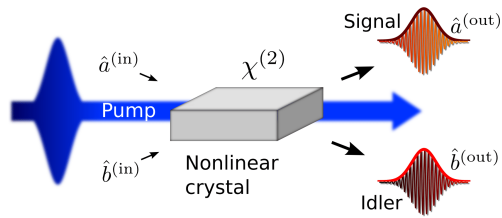}
\end{center}
\caption{Sketch of the type-II PDC process. A strong pump beam decays inside the nonlinear optical material into two beams usually labelled signal \(\hat{a}\) and idler \(\hat{b}\), forming a finitely squeezed EPR-state.}
\label{fig:parametric_downconversion_setup}
\end{figure}

The principle of type-II PDC is sketched in figure \ref{fig:parametric_downconversion_setup}. A strong pump beam decays inside a nonlinear optical material into two beams commonly labelled as signal and idler. In full generality, the generated output state created by type-II PDC is a finitely squeezed EPR-state or twin-beam state\footnote{Type-II PDC emits EPR states, whereas type-I PDC generates squeezed states. Type-I PDC is discussed in  \cite{wasilewski_pulsed_2006, lvovsky_decomposing_2007}.}.

As in the FC case, we do not restrict ourselves to a discussion of the type-II PDC process in the monochromatic picture, but extend the theories to the multi-mode picture including the interaction of many frequencies. This is especially important when the type-II PDC process is pumped by pulsed laser systems \cite{eckstein_highly_2011, gerrits_generation_2011}.

PDC in the high-gain regime has already been extensively studied: Wasilewski and Lvovsky investigated type-I PDC and its generation of squeezed states in ultrafast pulse modes in \cite{wasilewski_pulsed_2006, lvovsky_decomposing_2007}, while \cite{caspani_tailoring_2010} studies its spatio-temporal structure. Type-II PDC processes in the high-gain regime have been investigated as well \cite{brambilla_simultaneous_2004, gatti_multiphoton_2003, dayan_theory_2007}, yet with a focus on the correlations between the signal and idler beams. The theoretical framework for FC processes, presented in this paper, however, enables us to extend the works of Wasilwewski and Lvovsky \cite{wasilewski_pulsed_2006, lvovsky_decomposing_2007} on ultrafast type-I PDC to type-II PDC processes: We investigate the amount of generated EPR squeezing in the high-gain regime, the corresponding ultrafast mode shapes and the implications for experimental implementations.

\section{PDC: Hamiltonian}\label{sec:pdc_hamiltonian}
The interaction Hamiltonian of the type-II PDC process --- using electric field operators as defined in \eref{eq:electric_field_1D} and the corresponding approximations --- takes on the form
\begin{equation}
        \fl \qquad \qquad \hat{H}_I^{(PDC)}(t) = \epsilon_0 \, \int \mathrm dz \,  \chi^{(2)}(z) \, E_p^{(+)}(z,t)  \hat{E}_a^{(-)}(z,t) \hat{E}_b^{(-)}(z,t) + h.c. \,.
    \label{eq:pdc_interaction_hamiltonian}
\end{equation}
As in the FC process we assume a strong pump field exceeding the amplitudes of the signal and idler fields by several orders of magnitude and hence treat it as a classical field propagating undepleted through the medium (see \eref{eq:classical_pump_field}).

Using the free space propagation Hamiltonian from \eref{eq:fc_free_space_hamiltonian} and the interaction Hamiltonian from \eref{eq:pdc_interaction_hamiltonian} the process of type-II PDC is given by the overall Hamiltonian:
\begin{equation}
    \hat{H}_{PDC}(t) = \hat{H}_0^{(a)}(t) + \hat{H}_0^{(b)}(t) + \hat{H}_I^{(PDC)}(t)
    \label{eq:pdc_hamiltonian}
\end{equation}
While, at first glance, the process of type-II PDC seems very different from the process of FC, comparing the interaction Hamiltonian of PDC in \eref{eq:pdc_interaction_hamiltonian} and FC in \eref{eq:fc_interaction_hamiltonian} reveals that they both feature bilinear Hamitonians --- the pump is treated as a classical field --- with an almost identical structure and hence share many mathematical properties. 

It is therefore straightforward to extend our presented FC calculations to type-II PDC. In order to avoid repetition we are going to only state the results and elaborate on the differences and similarities to the process of FC. A detailed derivation is given in \cite{christ_theory_2013}.

\section{PDC: general properties}\label{sec:pdc_general_properties}
The general solution of \eref{eq:pdc_hamiltonian} takes on the form of a linear operator transformation \cite{ekert_relationship_1991, braunstein_squeezing_2005}:
\begin{eqnarray}
    \nonumber
    \fl \qquad \hat{a}^{\mathrm{(out)}}(\omega) = \int \mathrm d \omega' U_a(\omega, \omega')\, \hat{a}^{\mathrm{(in)}}(\omega') + \int \mathrm d \omega' V_a(\omega, \omega')\, \hat{b}^{\mathrm{(in)}\dagger}(\omega'),\\
    \fl \qquad \hat{b}^{\mathrm{(out)}}(\omega) = \int \mathrm d \omega' U_b(\omega, \omega')\, \hat{b}^{\mathrm{(in)}}(\omega') + \int \mathrm d \omega' V_b(\omega, \omega')\, \hat{a}^{\mathrm{(in)}\dagger}(\omega').
    \label{eq:pdc_general_solution}
\end{eqnarray}
This solution is constrained by the fact that it has to form a canonical transformation \cite{ekert_relationship_1991, braunstein_squeezing_2005, wasilewski_pulsed_2006}. Under this restriction we are able to rewrite it as
\begin{eqnarray}
    \nonumber
    &\hat{A}_k^{\mathrm{(out)}} = \cosh(r_k) \hat{A}_k^{\mathrm{(in)}} + \sinh(r_k) \hat{B}_k^{\mathrm{(in)}\dagger},\\
    &\hat{B}_k^{\mathrm{(out)}} = \cosh(r_k) \hat{B}_k^{\mathrm{(in)}} + \sinh(r_k) \hat{A}_k^{\mathrm{(in)}\dagger},
    \label{eq:pdc_general_broadband_solution}
\end{eqnarray}
where \(\hat{A}_k\) and \(\hat{B}_k\) are defined as broadband single-photon destruction operators \cite{rohde_spectral_2007}:
\begin{eqnarray}
    \nonumber
    \fl \qquad &\hat{A}_k^{\mathrm{(out)}} = \int \mathrm d \omega \, \varphi_k(\omega)\, \hat{a}^{\mathrm{(out)}}(\omega), \qquad
    \hat{B}_k^{\mathrm{(out)}} = \int \mathrm d \omega \, \xi_k(\omega) \, \hat{b}^{\mathrm{(out)}}(\omega), \\
    \fl \qquad  &\hat{A}_k^{\mathrm{(in)}} = \int \mathrm d \omega \, \psi_k(\omega) \,\hat{a}^{\mathrm{(in)}}(\omega), \qquad \,\,\,\,\,
    \hat{B}_k^{\mathrm{(in)}} = \int \mathrm d \omega \, \phi_k(\omega) \, \hat{b}^{\mathrm{(in)}}(\omega).
    \label{eq:pdc_broadband_mode_operators}
\end{eqnarray}
The details of this procedure are given in \ref{app:PDC_canonical_transformation_conditions}.

According to \eref{eq:pdc_general_broadband_solution} the type-II PDC process generates a number of finitely squeezed EPR-states \cite{barnett_methods_2003} generated in ultrafast optical pulse modes \(\hat{A}_k\) and \(\hat{B}_k\). The crucial parameters of this transformation are firstly the EPR amplitudes \(r_k\) which give both the amount of generated EPR squeezing --- \(\mathrm{EPR} \,\, \mathrm{squeezing[dB]} = -10 \log_{10}\left(e^{-2 r_k}\right)\) --- and the number of emitted EPR states, and secondly the mode shapes \(\varphi_k(\omega)\), \(\xi_k(\omega)\), \(\psi_k(\omega)\), and \(\phi_k(\omega)\), which define the shape in which the EPR states are emitted.

\section{PDC: analytic model excluding time-ordering effects}\label{sec:pdc_analytic_model_excluding_time-ordering_effects}
As in the FC case, presented in section \ref{sec:fc_analytic_model_excluding_time-ordering_effects}, we first solve the process excluding time-ordering effects.\footnote{The physical meaning of the time-ordering operator in type-II PDC is discussed in the works of Bra\'{n}czyk \cite{branczyk_non-classical_2010, branczyk_time_2011}.} Again we use the electric fields in the frequency domain \eref{eq:fc_electric_field_1D_frequency} and move into the interaction picture. Retracting the steps from section \ref{sec:fc_analytic_model_excluding_time-ordering_effects} we obtain
\begin{eqnarray}
    \fl \qquad \hat{U}_{PDC} = \exp\left[-\frac{\imath}{\hbar}\left(\int \mathrm d \omega_a \, \int \mathrm d \omega_b \, f(\omega_a, \omega_b) \hat{a}^\dagger(\omega_a) \hat{b}^\dagger(\omega_b) + h.c. \right) \right],
    \label{eq:pdc_no_time_ordering}
\end{eqnarray}
where \(f(\omega_a, \omega_b)\) is defined as
\begin{eqnarray}
    f(\omega_a, \omega_b) = B\, \alpha(\omega_a + \omega_b)  \, \mathrm{sinc}\left(\frac{\Delta k(\omega_a, \omega_b) L}{2}\right)
    \label{eq:pdc_joint_spectral_amplitude}
\end{eqnarray}
and \(\Delta k(\omega_a, \omega_b) = k_p(\omega_a + \omega_b) - k_a(\omega_a) - k_b(\omega_b)\).
Again using the broadband mode formalism, we are able to write \(\hat{U}_{PDC}\) in the Heisenberg formalism. It takes on the form
\begin{eqnarray}
    \nonumber
    &\hat{A}_k^{\mathrm{(out)}} = \cosh(r_k) \hat{A}_k^{\mathrm{(in)}} + \sinh(r_k) \hat{B}_k^{^\dagger\mathrm{(in)}},\\
    &\hat{B}_k^{\mathrm{(out)}} = \cosh(r_k) \hat{B}_k^{\mathrm{(in)}} + \sinh(r_k) \hat{A}_k^{^\dagger\mathrm{(in)}}.
    \label{eq:pdc_no_time_ordering_broadband_input_output_relation}
\end{eqnarray}
The details of this calculation are given in \cite{christ_probing_2011}.

This result exhibits exactly the structure imposed by the canonical commutation relation in \eref{eq:pdc_general_broadband_solution}, except for the fact that, as in the FC case, the input and output modes are of identical shape.

In conclusion, the analytic model ignoring time-ordering effects enables a straightforward solution of the type-II PDC process, which enables the efficient engineering and design of type-II PDC processes as long as the applied approximations hold.

\section{PDC: rigorous theory including time-ordering effects}\label{sec:pdc_rigorous_theory_including_time-ordering_effects}
Having elaborated on solving type-II PDC neglecting time-ordering effects, we further built a rigorous model of the process. For this purpose we adapt the approach presented in section \ref{sec:fc_rigorous_theory_including_time-ordering_effects}.

Repeating exactly the same steps as in section \ref{sec:fc_rigorous_theory_including_time-ordering_effects}, we obtain two operator-valued integro-differential equations describing the type-II PDC process:
\begin{eqnarray}
    \nonumber
    & \frac{\partial}{\partial z}\hat{\epsilon}_a(z,\omega)= \int \mathrm d\omega' f(\omega, \omega', z) \, \hat{\epsilon}_b^\dagger(z,\omega'),\\
    & \frac{\partial}{\partial z}\hat{\epsilon}_b(z,\omega)= \int \mathrm d\omega' f(\omega', \omega, z) \,  \hat{\epsilon}_a^\dagger(z,\omega') 
    \label{eq:pdc_diff_eq_final}
\end{eqnarray}
with
\begin{equation}
    f(\omega, \omega', z) = - \frac{\imath}{\hbar} D E_p^{(+)}(z, \omega + \omega') \exp\left[\imath \Delta k(\omega, \omega') z \right].
    \label{eq:pdc_f_function}
\end{equation}
Here we introduced the shorthand \( \Delta k(\omega, \omega') =  k_p(\omega + \omega') - k_a(\omega) - k_b(\omega') \). The structure of this result is very similar to the equations derived by \cite{brambilla_simultaneous_2004, mauerer_colours_2008, wasilewski_pulsed_2006}, which serves as a nice cross check of our calculations. Also take note of the switch of \(\omega\) and \(\omega'\) in \(f\) in the two equations in \eref{eq:pdc_diff_eq_final}.

\subsection{Solving the differential equations}\label{sec:pdc_solving_the_diff_eq}
Since the structure of the two differential equations describing the type-II PDC process in \eref{eq:pdc_diff_eq_final} is identical to those describing the FC process in \eref{eq:fc_diff_eq_final_1} and \eref{eq:fc_diff_eq_final_2}, we apply the same solution method as presented in section \ref{sec:fc_solving_the_diff_eq}.

We obtain four classical integro-differential equations. Two for \(U_a(z, \omega, \omega')\) and \(V_b(z, \omega, \omega')\):
\begin{eqnarray}
    \nonumber
    &\frac{\partial}{\partial z}U_a(z,\omega,\omega'') = \int \mathrm d\omega' f(\omega, \omega', z)  V_b^*(z,\omega',\omega''),\\
    & \frac{\partial}{\partial z}V_b(z,\omega,\omega'') = \int \mathrm d\omega' f(\omega', \omega, z)  U_a^*(z,\omega',\omega'').
    \label{eq:pdc_classical_differential_equation_1}
\end{eqnarray}
And two for \(U_b(z, \omega, \omega')\) and \(V_a(z, \omega, \omega')\):
\begin{eqnarray}
    \nonumber
    & \frac{\partial}{\partial z}U_b(z,\omega,\omega'') = \int \mathrm d\omega' f(\omega', \omega, z)  V_a^*(z,\omega',\omega''),\\
    & \frac{\partial}{\partial z}V_a(z,\omega,\omega'')
    = \int \mathrm d\omega' f(\omega, \omega', z)  U_b^*(z,\omega',\omega'').
    \label{eq:pdc_classical_differential_equation_2}
\end{eqnarray}
The initial conditions are:
\begin{eqnarray}
    \nonumber
    U_a(z, \omega, \omega'') = U_b(\omega, \omega'', z) = \delta(\omega - \omega''),\\
    V_a(z, \omega, \omega'') = V_b(\omega, \omega'', z) = 0.
    \label{eq:pdc_initial_conditions}
\end{eqnarray}
These classical integro-differential equations are very similar to the ones derived by Brambilla in \cite{brambilla_simultaneous_2004}. As in the FC case, they can be solved via an iterative approach. Details of this calculation and the numerical errors in the solution method are give in \ref{app:pdc_numerical_implementation}. The program code, written in Python, is available, together with the FC code, on our website.

\section{PDC: comparison between simplified analytical and rigorous approach}\label{sec:pdc_comparison_between_simplified_analytical_and_rigorous_approach}
As in the FC case, presented in section \ref{sec:fc_comparison_between_simplified_analytical_and_rigorous_approach}, we consider an almost uncorrelated process pumped by ultrafast pump lasers, to compare the different approaches  presented in sections \ref{sec:pdc_analytic_model_excluding_time-ordering_effects} and  \ref{sec:pdc_rigorous_theory_including_time-ordering_effects}, since this is the case where the differences are most prominent. The process properties are given in \ref{app:pdc_simulated_down-conversion_process}, whereas the numerical implementation is detailed in  \ref{app:pdc_numerical_implementation}.

\begin{figure}[htb]
    \begin{center}
        \includegraphics[width=\textwidth]{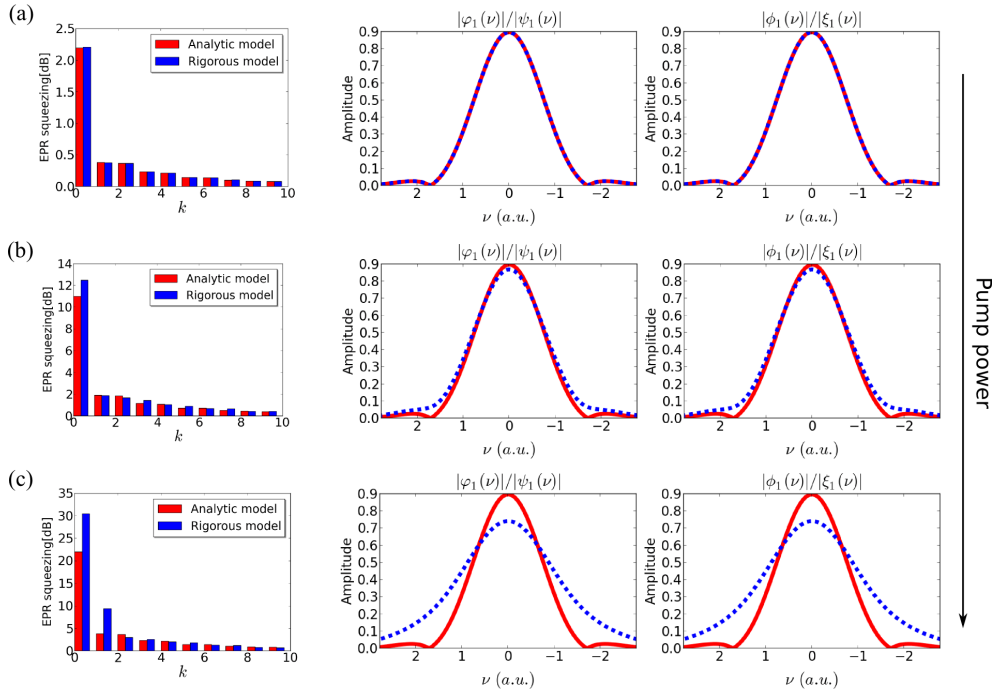}
    \end{center}
    \caption{Comparison between the rigorous and the analytical approach in (almost) uncorrelated few-mode type-II PDC. For low down-conversion rates, presented in (a) (\(\langle n \rangle = 0.07/0.07\) in the analytic/rigorous model), both approaches evaluate to identical results. Only in the case of rising EPR squeezing values in (b) (\(\langle n \rangle = 2.80/4.08\) in the analytic/rigorous model), with EPR squeezing values about 12dB, the two approaches start to show minor differences, which become more prominent when even higher EPR squeezing values are considered (c) (\(\langle n \rangle = 39.39/279.87\) in the analytic/rigorous model). (The process parameters are given in \ref{app:pdc_simulated_down-conversion_process}.)}
    \label{fig:pdc_time-ordering_result_uncorrelated}
\end{figure}

This analysis yielded the individual mode functions and corresponding \(r_k\)-values. While the \(r_k\)-parameters are, in principle, sufficient to describe the optical gain, we further evaluated the corresponding EPR squeezing values --- obtained via the relation \(\mathrm{squeezing[dB]} = -10 \log_{10}\left(e^{-2 r_k}\right)\) --- together with mean photon numbers --- \(\langle n \rangle = \sum_k \sinh^2(r_k)\) --- to ease the comparison with experimental implementations. We depicted the obtained EPR squeezing values together with the mode shapes in figure \ref{fig:pdc_time-ordering_result_uncorrelated} for rising pump powers from (a) to (c). The corresponding mean photon numbers are stated in the corresponding figure caption. The figures in the column on the left shows the EPR squeezing and the corresponding mean photon number of the complete state, whereas the two columns on the right present the corresponding mode functions \(\varphi_1(\nu), \psi_1(\nu), \phi(\nu)\) and \(\xi_1(\nu)\) for the first optical mode, where \(\varphi_1(\nu)\) and  \(\psi_1(\nu)\) as well as \(\phi_1(\nu)\) and \(\xi_1(\nu)\) are of identical shape. ``Analytic model'' labels the solution excluding time-ordering effects, as presented in section \ref{sec:pdc_analytic_model_excluding_time-ordering_effects} and the ``Rigorous model'' label marks the rigorous solution from section \ref{sec:pdc_rigorous_theory_including_time-ordering_effects}.

Up to EPR squeezing values of 12dB, corresponding to mean photon numbers about \(\langle n \rangle = 4\), presented in figure \ref{fig:pdc_time-ordering_result_uncorrelated} (b) the two approaches give identical results, only when EPR squeezing values beyond this bound are considered significant differences between the two models start to appear. The rigorous model predicts more EPR squeezing than the analytic model and the time-ordering leads to a broadening of the mode shapes in the high-gain regime.

\section{PDC: implications for experimental implementations}\label{sec:pdc_implications}
In summary we expanded the works of Wasilewski and Lvovsky \cite{wasilewski_pulsed_2006, lvovsky_decomposing_2007} to type-II PDC. The analytical model accurately describes type-II PDC in the low-gain regime up to EPR squeezing values of 12dB, where minor deviations start to appear. Only for extremely high EPR squeezing values, in the range of 20dB and higher, complicated non-trivial deviations from the analytical model appear and give significant contributions that require a rigorous treatment of type-II PDC. For most experimental setups and applications, it is hence perfectly justified to apply the simplified analytic model to minimize the computational effort, as long as its limitations are kept in mind.

To date type-II PDC processes are mainly used for three purposes: The generation of EPR-states \cite{eckstein_highly_2011, gerrits_generation_2011}, entangled photon-pair generation \cite{kwiat_new_1995, kwiat_ultrabright_1999, kurtsiefer_high-efficiency_2001}, and the heralding of single-photons \cite{uren_efficient_2004, pittman_heralding_2005, castelletto_optimizing_2006}, especially pure single-photons \cite{mosley_heralded_2008, poh_eliminating_2009, evans_bright_2010}.

Considering experiments aiming to generate EPR-states in ultrafast optical modes \cite{eckstein_highly_2011, gerrits_generation_2011} time-ordering effects have to be taken into account as soon as EPR squeezing values exceeding 12dB are considered. In contrast to FC processes, however, the time-ordering effects are beneficial to the performance of the sources. They lead to much higher EPR squeezing values as predicted by the simplified analytic model.

When the generation of entangled photon-pairs is considered, higher-order photon-pair contributions are detrimental to the performance of the source; it is hence necessary to pump these type-II PDC processes with the lowest pump power available. In this regime, time-ordering effects can be neglected and the analytical model is fully sufficient to investigate the impacts of higher-order photon-number effects on the quality of the generated entanglement.

In the case of heralding single-photons from type-II PDC, the first theoretical descriptions restricted themselves to a description of type-II PDC using first-order perturbation theory \cite{grice_spectral_1997}. With the brightness of PDC sources steadily increasing attention has fallen on the effects of higher-order photon numbers on the purity of the heralded states \cite{osullivan_conditional_2008, huang_optimized_2011, wasilewski_statistics_2008}. For the heralding of single-photons from PDC, however, it has been shown that the required mean photon numbers for optimal rates and purity are about one \(\langle n \rangle = 1\) \cite{christ_limits_2012}. In this regime our simplified analytic model, presented in section \ref{sec:pdc_analytic_model_excluding_time-ordering_effects}, is sufficient to appropriately model the process. This is very beneficial for the engineering of advanced single-photon sources, since this enables quick and straightforward analytic calculations, which greatly enhances the engineering process.

\section{Conclusion}\label{sec:conclusion}
In conclusion, we developed two models for the nonlinear optical processes of FC and type-II PDC, taking into account higher-order photon number effects. The presented rigorous numerical model relies on the solution of coupled differential equations, whereas ignoring time-ordering effects enabled us to construct an analytical solution.

Our analysis revealed that the presented analytic model gives accurate results for many experimental configurations: In the case of FC processes below unit conversion efficiency the analytic model is sufficient. At unit conversion efficiency, however, the rigorous model has to be applied, which predicts a significant decrease in the performance of quantum pulse gate applications. Type-II PDC is accurately described by the analytic model up to EPR squeezing values of 12dB, which is sufficient to model type-II PDC in entangled photon-pair generation and single-photon heralding experiments. Above the 12dB bound however the rigorous approach has to be applied, which predicts an enhanced EPR squeezing generation rate.

\section{Acknowledgments}
The authors thank Agata M. Bra\'{n}czyk and Regina Kruse for useful discussions and helpful comments.

The research leading to these results has received funding from the European Community’s Seventh Framework Programme FP7/2007-2013 under the grant agreement Q-Essence 248095.

\begin{appendix}
    
\section{FC: canonical transformation conditions}\label{app:FC_canonical_transformation_conditions}
The FC process in \eref{eq:fc_general_solution} is a unitary process. In the Heisenberg picture, we are able to write the FC unitary \(\hat{U}_{FC}\) as an operator transformation. For the operator transformation to represent a unitary process, the transformation has to preserve the canonical commutation relations. This imposes several restrictions on the structure of the solution. We evaluate these by extending the calculations from \cite{ekert_relationship_1991, braunstein_squeezing_2005} to FC. At first we rewrite \eref{eq:fc_general_solution} in the more compact notation
\begin{eqnarray}
    \nonumber
    \hat{a}_i^{\mathrm{(out)}} = u_{ij}^{(a)}\hat{a}_j^{\mathrm{(in)}} + v_{ij}^{(a)} \hat{c}_j^{\mathrm{(in)}},\\
    \hat{c}_i^{\mathrm{(out)}} = u_{ij}^{(c)}\hat{c}_j^{\mathrm{(in)}} - v_{ij}^{(c)} \hat{a}_j^{\mathrm{(in)}},
    \label{eq:fc_general_solution_shorthand}
\end{eqnarray}
where \(i\) and \(j\) label the individual frequencies of the electric fields and summation over repeated indices is understood. These two input-output relations must preserve
\begin{eqnarray}
    \nonumber
    \left[\hat{a}_i, \hat{a}_j^{\dagger}\right] = \left[\hat{c}_i, \hat{c}_j^{\dagger}\right] = \delta_{ij},\\
    \left[\hat{a}_i, \hat{c}_j^{\dagger}\right] = 0.
    \label{eq:fc_commutation_relations}
\end{eqnarray}
Using \eref{eq:fc_general_solution_shorthand} and \eref{eq:fc_commutation_relations} we obtain three conditions for FC:
\begin{eqnarray}
    U_a U_a^\dagger + V_a V_a^\dagger \, = \, U_c U_c^\dagger + V_c V_c^\dagger \, = \, I,
    \label{eq:fc_canonical_conditions_id}\\
    U_a V_c^\dagger - V_a U_c^\dagger \, = \, 0.
    \label{eq:fc_canonical_conditions_zero}
\end{eqnarray}
Furthermore the commutation relations have to be preserved for the inverse transformation as well:
\begin{eqnarray}
    \nonumber
    \hat{a}_i^{\mathrm{(in)}} = u_{ji}^{*(a)}\hat{a}_j^{\mathrm{(out)}} - v_{ji}^{*(c)} \hat{c}_j^{\mathrm{(out)}},\\
    \hat{c}_i^{\mathrm{(in)}} = u_{ji}^{*(c)}\hat{c}_j^{\mathrm{(out)}} + v_{ji}^{*(a)} \hat{a}_j^{\mathrm{(out)}}.
    \label{eq:fc_general_inv_solution_shorthand}
\end{eqnarray}
With \eref{eq:fc_general_inv_solution_shorthand} the canonical commutation conditions in \eref{eq:fc_commutation_relations} yield the restrictions:
\begin{eqnarray}
    U_a^\dagger U_a + V_c^\dagger V_c \, = \, U_c^\dagger U_c + V_a^\dagger V_a \, = \, I,
    \label{eq:fc_inv_canonical_conditions_id}\\
    U_a^\dagger V_a - V_c^\dagger U_c \, =  \, 0.
    \label{eq:fc_inv_canonical_conditions_zero}
\end{eqnarray}
The equations \eref{eq:fc_canonical_conditions_id}, \eref{eq:fc_canonical_conditions_zero}, \eref{eq:fc_inv_canonical_conditions_id} and \eref{eq:fc_inv_canonical_conditions_zero} impose several constraints on the solution. However, they are rather unintuitive representations of the symmetries governing the FC process, yet with the help of the Bloch-Messiah reduction \cite{braunstein_squeezing_2005} it is possible to unravel their underlying structure: As a first step we decompose the four matrices \(U_a, V_a, U_c, V_c\) as
\begin{eqnarray}
    \nonumber
    U_a = A_a^u D_a^u B_a^{u\dagger}, \qquad V_a = A_a^v D_a^v B_a^{v\dagger},\\
    U_c = A_c^u D_c^u B_c^{u\dagger}, \qquad V_c = A_c^v D_c^v B_c^{v\dagger}.
    \label{eq:fc_svd_compact}
\end{eqnarray}
where \(A\) and \(B\) are unitary matrices and \(D\) is a diagonal matrix with real entries. This definition is equivalent to a SVD except for the fact that we allow the individual elements in \(D\) to exhibit negative values.\footnote{The reason for this extension of the SVD becomes clear in \eref{eq:broadband_fc_beamspitter_transformation}.}

The matrices \(U_a U_a^\dagger\) and \(V_a V_a^\dagger\) are Hermitian and \eref{eq:fc_canonical_conditions_id}, implies that they commute; hence both are diagonalised by the same unitary matrix \(P\):
\begin{eqnarray}
    P\, U_a U_a^\dagger P^\dagger = D, \qquad P\, V_a V_a^\dagger P^\dagger = D'.
    \label{eq:fc_P_diagonalization}
\end{eqnarray}
With the help of the decomposition in \eref{eq:fc_svd_compact} they can be written as
\begin{eqnarray}
    P A_a^u D_a^{u^2} A_a^{u^\dagger} P^\dagger = D, \qquad P A_a^v D_a^{v^2} A_a^{v^\dagger} P^\dagger = D'.
\end{eqnarray}
And we obtain \(A_a^u = A_a^v\). From \eref{eq:fc_canonical_conditions_id} using \(U_c U_c^\dagger\) and \(V_c V_c^\dagger\) we infer in a similar manner \(A_c^u = A_c^v\). Evaluating the conditions for the inverse transformation using \eref{eq:fc_inv_canonical_conditions_id} yields \(B_a^u = B_c^v\) and \(B_c^u = B_a^v\). Consequently the decomposition in \eref{eq:fc_svd_compact} can be written as:
\begin{eqnarray}
    \nonumber
    &U_a = A_a D_a^u B_a^\dagger, \qquad U_c = A_c D_c^u B_c^\dagger, \\
    &V_a = A_a D_a^v B_c^\dagger, \qquad V_c = A_c D_c^v B_a^\dagger.
\label{eq:fc_svd_simplified_1}
\end{eqnarray}
Using the matrices in  \eref{eq:fc_svd_simplified_1} in conjunction with the conditions in \eref{eq:fc_canonical_conditions_id} we further obtain:
\begin{equation}
    D_a^{u^2} + D_a^{v^2} = I, \qquad D_c^{u^2} + D_c^{v^2} = I.
\end{equation}
This implies that the individual elements of the \(D\) matrices have to obey \(\cos(r_k)\) and \(\sin(r_k)\) behaviour. Applying the conditions in \eref{eq:fc_inv_canonical_conditions_id} to the transformation matrices in \eref{eq:fc_svd_simplified_1} results in
\begin{equation}
    D_a^{u^2} + D_c^{v^2} = I, \qquad D_c^{u^2} + D_a^{v^2} = I,
\end{equation}
from which we conclude \(D_a^{u^2} = D_c^{u^2} = D^{u^2}\) and \(D_a^{v^2} = D_c^{v^2} = D^{v^2}\). Taking everything into account, the final decomposed FC matrices read
\begin{eqnarray}
    \nonumber
    &U_a = A_a D^u B_a^\dagger, \qquad U_c = A_c D^u B_c^\dagger, \\
    \nonumber
    &V_a = A_a D^v B_c^\dagger, \qquad V_c = A_c D^v B_a^\dagger,  \\
    &\qquad \qquad D^{u^2} + D^{v^2} = I.
\label{eq:fc_svd_simplified_2}
\end{eqnarray}
In the original representation we consequently require
\begin{eqnarray}
    \nonumber
    &U_a(\omega, \omega') = \sum_k \varphi_k^*(\omega) \cos(r_k) \psi_k(\omega'), \\
    \nonumber
    &V_a(\omega, \omega') = \sum_k \varphi_k^*(\omega) \sin(r_k) \phi_k(\omega'), \\
    \nonumber
    &U_c(\omega, \omega') = \sum_k \xi_k^*(\omega) \cos(r_k) \phi_k(\omega'), \\
    &V_c(\omega, \omega') = \sum_k \xi_k^*(\omega) \sin(r_k) \psi_k(\omega').
    \label{eq:fc_svd_symmetries}
\end{eqnarray}
From these symmetry relations the FC process in \eref{eq:fc_general_solution} must, in the Heisenberg picture, form a multitude of beam-splitter relations in orthogonal optical modes
\begin{eqnarray}
    \nonumber
    &\hat{A}_k^{\mathrm{(out)}} = \cos(r_k) \hat{A}_k^{\mathrm{(in)}} + \sin(r_k) \hat{C}_k^{\mathrm{(in)}},\\
    &\hat{C}_k^{\mathrm{(out)}} = \cos(r_k) \hat{C}_k^{\mathrm{(in)}} - \sin(r_k) \hat{A}_k^{\mathrm{(in)}}, 
    \label{eq:broadband_fc_beamspitter_transformation}
\end{eqnarray}
where we defined:
\begin{eqnarray}
    \nonumber
    \fl \qquad &\hat{A}_k^{\mathrm{(out)}} = \int \mathrm d \omega \, \varphi_k(\omega) \hat{a}^{\mathrm{(out)}}(\omega), \qquad
    \hat{C}_k^{\mathrm{(out)}} = \int \mathrm d \omega \, \xi_k(\omega) \hat{c}^{\mathrm{(out)}}(\omega), \\
    \fl \qquad  &\hat{A}_k^{\mathrm{(in)}} = \int \mathrm d \omega \, \psi_k(\omega) \hat{a}^{\mathrm{(in)}}(\omega), \qquad \,\,\,\,\,
    \hat{C}_k^{\mathrm{(in)}} = \int \mathrm d \omega \, \phi_k(\omega) \hat{c}^{\mathrm{(in)}}(\omega).
\end{eqnarray}
Note however that the canonical commutation relations do not demand that the input and output modes are of identical shape. In principle the input modes \(\hat{A}^{\mathrm{(in)}}\) and output modes \(\hat{A}^{\mathrm{(out)}}\) could feature completely different spectral mode functions \(\varphi_k(\omega)\) and \(\psi_k(\omega)\) but still form a canonical and hence unitary solution.

\section{PDC: canonical transformation conditions}\label{app:PDC_canonical_transformation_conditions}
As in the case of FC the type-II PDC process is described by a unitary transformation, hence it must preserve the canonical commutation relations. Retracting the calculation in \ref{app:FC_canonical_transformation_conditions} and adapting the work from \cite{ekert_relationship_1991, braunstein_squeezing_2005} to type-II PDC, they read:
\begin{eqnarray}
    U_a U_a^\dagger - V_a V_a^\dagger \, = \, U_b U_b^\dagger - V_b V_b^\dagger \, = \, I, 
    \label{eq:pdc_canonical_conditions_id}\\
    U_a V_b^T - V_a U_b^T \, = \, 0.
    \label{eq:pdc_canonical_conditions_zero}
\end{eqnarray}
For the inverse transformation they evaluate to
\begin{eqnarray}
    U_a^\dagger U_a - (V_b^\dagger V_b)^T \, = \, U_b^\dagger U_b - (V_a^\dagger V_a)^T \, = \, I,
    \label{eq:pdc_inv_canonical_conditions_id}\\
    U_a^\dagger V_a - (U_b^\dagger V_b)^T \, =  \, 0.
    \label{eq:pdc_inv_canonical_conditions_zero}
\end{eqnarray}
With the help of the SVD theorem and \eref{eq:pdc_canonical_conditions_id}, \eref{eq:pdc_canonical_conditions_zero} and \eref{eq:pdc_inv_canonical_conditions_zero} the four matrices of a general type-II PDC process in \eref{eq:pdc_general_solution} are restricted to the form
\begin{eqnarray}
    \nonumber
    &U_a(\omega, \omega') = \sum_k \varphi_k^*(\omega) \cosh(r_k) \psi_k(\omega'), \\
    \nonumber
    &V_a(\omega, \omega') = \sum_k \varphi_k^*(\omega) \sinh(r_k) \phi_k^*(\omega'), \\
    \nonumber
    &U_b(\omega, \omega') = \sum_k \xi_k^*(\omega) \cosh(r_k) \phi_k(\omega'), \\
    &V_b(\omega, \omega') = \sum_k \xi_k^*(\omega) \sinh(r_k) \psi_k^*(\omega').
    \label{eq:pdc_svd_symmetries}
\end{eqnarray}
From these symmetry relations, the type-II PDC process \eref{eq:pdc_general_solution} consists of multiple twin-beam squeezers in orthogonal optical modes:\footnote{In principle the twin-beam squeezer has a phase degree of freedom \cite{barnett_methods_2003}, which we absorb in the definition of the electric field operators \(\hat{A}_k\) and \(\hat{B}_k\).}
\begin{eqnarray}
    \nonumber
    &\hat{A}_k^{\mathrm{(out)}} = \cosh(r_k) \hat{A}_k^{\mathrm{(in)}} + \sinh(r_k) \hat{B}_k^{\mathrm{(in)}\dagger}\\
    &\hat{B}_k^{\mathrm{(out)}} = \cosh(r_k) \hat{B}_k^{\mathrm{(in)}} + \sinh(r_k) \hat{A}_k^{\mathrm{(in)}\dagger} 
    \label{eq:broadband_pdc_beamspitter_transformation}
\end{eqnarray}
where we defined:
\begin{eqnarray}
    \nonumber
    \fl \qquad &\hat{A}_k^{\mathrm{(out)}} = \int \mathrm d \omega \, \varphi_k(\omega) \hat{a}^{\mathrm{(out)}}(\omega) \qquad
    \hat{B}_k^{\mathrm{(out)}} = \int \mathrm d \omega \, \xi_k(\omega) \hat{b}^{\mathrm{(out)}}(\omega) \\
    \fl \qquad  &\hat{A}_k^{\mathrm{(in)}} = \int \mathrm d \omega \, \psi_k(\omega) \hat{a}^{\mathrm{(in)}}(\omega) \qquad \,\,\,\,\,
    \hat{B}_k^{\mathrm{(in)}} = \int \mathrm d \omega \, \phi_k(\omega) \hat{b}^{\mathrm{(in)}}(\omega)
\end{eqnarray}
Note however that, as in the FC case, the canonical commutation relations do not demand that the input and output modes are of identical shape.

\section{FC: simulated FC processes}\label{app:fc_simulated_frequency_conversion_process}
In our simulation of FC, we did not restrict ourselves to a specific crystal material or wavelength range, but created a generic model of the process. For this purpose we first moved from the \( (\omega, \omega') \)-system to the parameter range \( (\nu, \nu') \) relative to the central frequencies of the FC process \( (\omega_{0}, \omega'_{0}) \). In the simulation we work with a Gaussian pump distribution, as created by pulsed laser systems. The pump distribution in \eref{eq:fc_joint_spectral_amplitude} and \eref{eq:fc_diff_eq_final_1} takes on the form
\begin{equation}
    \alpha(\nu-\nu') = E_p \exp\left[ -\frac{(\nu-\nu')^2}{2 \sigma^2}\right], 
\end{equation}
where \(E_p\) labels the pump amplitude and \(\sigma\) the pump width. The second function we have to adapt is the phasematching function \(\Delta k(\omega, \omega') = k_p(\omega' - \omega) - k_c(\omega') + k_a(\omega)\). As a first step we perform a Taylor expansion of the individual \(k(\omega)\) terms up to first order about their central frequency \(\omega_0\):
\begin{eqnarray}
    k(\omega) &\approx k(\omega_0) + \frac{d}{d \omega} k(\omega_0) \underbrace{(\omega - \omega_0)}_{\nu}.
\end{eqnarray}
This is justified since we restrict ourselves to nonlinear processes not to broad in frequency (slowly varying envelope approximation \(\Delta \omega \ll \omega_0\)) far from any singularities in the dispersion relation. At the central frequencies the process, per definition, displays perfect phasematching \( k_p(\omega'_0 - \omega_0) - k_c(\omega'_0) + k_a(\omega_0) = 0 \) and the phasematching function simplifies to
\begin{eqnarray}
    \fl \qquad \qquad \Delta k(\nu, \nu') = \frac{d}{d \omega}k_p(\omega'_0 - \omega_0) (\nu' - \nu) - \frac{d}{d \omega}k_c(\omega_0') \, \nu' + \frac{d}{d \omega} k_a(\omega_0) \, \nu.
\end{eqnarray}
The three remaining parameters \(\frac{d}{d \omega}k_p(\omega'_0 - \omega_0)\), \(\frac{d}{d \omega}k_c(\omega_0')\) and \(\frac{d}{d \omega}k_a(\omega_0)\) --- the inverse group velocities of the three interacting beams --- define the material properties of the system and can be adjusted accordingly. 

This compact notation enables us to simulate any FC process with the help of just six parameters. The width and amplitude of the pump beam, the group velocities of the three interacting waves and the length of the nonlinear medium. 

In order to evaluate (almost) uncorrelated FC with few optical modes \(r_k\), as depicted in figure \ref{fig:fc_time-ordering_result_uncorrelated}, we applied \(\sigma = 0.98190\), \(\frac{d}{d \omega}k_p(\omega'_0 - \omega_0) = 3.0 \), \(\frac{d}{d \omega}k_c(\omega_0') = 1.5 \), \(\frac{d}{d \omega}k_a(\omega_0) = 4.5\) and a crystal of length \(L = 2\). The pump amplitude \(E_p\) is adjusted to give the desired conversion rates.

\section{FC: numerical implementation}\label{app:fc_numerical_details}
In order to obtain the time-ordered solutions we solved the classical differential equations in \eref{eq:classical_differential_equations_1} and \eref{eq:classical_differential_equations_2} which give the functions \(U_a(z, \omega, \omega'')\), \(U_c(z, \omega, \omega'')\), \(V_a(z, \omega, \omega'')\), and \(V_c(z, \omega, \omega'')\) describing the FC process.

In the numerical implementation of FC we used a sampling of 500 points for each frequency degree of freedom and 500 points in \(z\)-direction to discretize the functions \(U_a(z, \omega, \omega'')\), \(U_c(z, \omega, \omega'')\), \(V_a(z, \omega, \omega'')\), \(V_c(z, \omega, \omega'')\) and \(f(\omega, \omega', z)\). We evaluated the successive integrations in \eref{eq:integrated_classical_differential_equations_1} via the trapezoid rule until the solutions converged. The actual solution defining the overall process properties is given by the matrices at the end of the crystal \(U_a(z = \frac{L}{2}, \omega, \omega'')\), \(U_c(z = \frac{L}{2}, \omega, \omega'')\), \(V_a(z = \frac{L}{2}, \omega, \omega'')\), and \(V_c(z = \frac{L}{2}, \omega, \omega'')\)

We checked the accuracy of the result in a variety of ways: At first we evaluated the canonical transformation conditions in \eref{eq:fc_canonical_conditions_id}, \eref{eq:fc_canonical_conditions_zero}, \eref{eq:fc_inv_canonical_conditions_id}, and \eref{eq:fc_inv_canonical_conditions_zero}. For example in the case of \eref{eq:fc_canonical_conditions_zero} we calculated:
\begin{eqnarray}
    \nonumber
    \fl \qquad \int \mathrm d \omega' U_a(z = \frac{L}{2}, \omega, \omega') V_c(z = \frac{L}{2}, \omega'', \omega')^* \\
    \nonumber
    - \int \mathrm d \omega' V_a(z = \frac{L}{2}, \omega, \omega') U_c(z = \frac{L}{2}, \omega'', \omega')^* \\
    = D^{(diff)}(z = \frac{L}{2}, \omega, \omega')
\end{eqnarray}
and determined the distance of \(D^{(diff)}(z = \frac{L}{2}, \omega, \omega')\) from the expected zero matrix and consequently the error in the solution via:
\begin{equation}
    \fl \qquad \mathrm{error} = \frac{\int \mathrm d \omega \int \mathrm d \omega' D^{(diff)}(z = \frac{L}{2},\omega, \omega')}{0.5 \left[ \int \mathrm d \omega \int \mathrm d \omega' |V_a(z = \frac{L}{2},\omega, \omega')| + \int \mathrm d \omega \int \mathrm d \omega' |U_c(z = \frac{L}{2},\omega, \omega')|\right]}
\end{equation}
In all presented cases the obtained \(\mathrm{error}\) was below 0.00027.

We also checked the numerical Schmidt decompositions of \(U_a(z = \frac{L}{2}, \omega, \omega'')\), \(U_c(z = \frac{L}{2}, \omega, \omega'')\), \(V_a(z = \frac{L}{2}, \omega, \omega'')\), and \(V_c(z = \frac{L}{2}, \omega, \omega'')\) to verify the mode properties derived in \ref{app:FC_canonical_transformation_conditions}. During this process the decompositions of \(U_a(z = \frac{L}{2}, \omega, \omega'')\), \(U_c(z = \frac{L}{2}, \omega, \omega'')\) showed numerical issues, these however could be resolved by decomposing \(U_a(z = \frac{L}{2}, \omega, \omega'') U_a^\dagger(z = \frac{L}{2}, \omega, \omega'')\), \(U_a^\dagger(z = \frac{L}{2}, \omega, \omega'') U_a(z = \frac{L}{2}, \omega, \omega'')\), \(U_c(z = \frac{L}{2}, \omega, \omega'') U_c^\dagger(z = \frac{L}{2}, \omega, \omega'')\), \(U_c^\dagger(z = \frac{L}{2}, \omega, \omega'') U_c(z = \frac{L}{2}, \omega, \omega'')\) instead. The obtained modes from these four matrices provided a much improved stability especially in the high-gain regime. Using these Schmidt modes we verified that the obtained Schmidt values of the \(U\) and \(V\) matrices behaved like \(\cos(r_k)^2 + \sin(r_k)^2 = 1\) with errors below 0.0001. We also asserted that the decompositions yielded the functions \(\varphi_k, \psi_k, \phi_k, \xi_k\) with symmetries as detailed in \eref{eq:fc_svd_symmetries}, which were fulfilled within numerical accuracy.

The program code, written in Python, published on our website, is able to directly create the investigated FC processes and also performs all mentioned tests. It further enables the simulation of actual frequency conversion processes. In this case the unit for the length of the crystal has to match with the inverse group velocities and the unit for the width of the pump beam with the applied units for \(\nu\) and \(\nu'\).

\section{PDC: simulated PDC process}\label{app:pdc_simulated_down-conversion_process}
As in the simulation of FC processes in \ref{app:fc_simulated_frequency_conversion_process} we didn't restrict ourselves to a specific crystal material and wavelength range but created a generic model of the process. Again we first move from the \( (\omega, \omega') \)-system to the parameter range \( (\nu, \nu') \) relative to the central frequencies of the type-II PDC process \( (\omega_0, \omega_0') \). As in the FC case we used a Gaussian pump distribution for the simulation, which in  \eref{eq:pdc_joint_spectral_amplitude} and \eref{eq:pdc_f_function} is given by
\begin{equation}
    \alpha(\nu+\nu') = E_p \exp\left[ -\frac{(\nu+\nu')^2}{2 \sigma^2}\right],
\end{equation}
where \(E_p\) labels the pump amplitude and \(\sigma\) the pump width. The second function we have to adapt is the phase-matching function \( \Delta k(\omega, \omega') = k_p(\omega + \omega') - k_a(\omega) - k_b(\omega') \). As a first step we perform a Taylor expansion of the individual \(k(\omega)\) terms up to first order about their central frequency \(\omega_0\):
\begin{eqnarray}
    k(\omega) &\approx k(\omega_0) + \frac{d}{d \omega} k(\omega_0) \underbrace{(\omega - \omega_0)}_{\nu}.
\end{eqnarray}
This is justified since we restrict ourselves to nonlinear processes not to broad in frequency (slowly varying envelope approximation \(\Delta \omega \ll \omega_0\)) far from any singularities in the dispersion. At the central frequencies the process, per definition, displays perfect phase-matching \( k_p(\omega'_0 + \omega_0) - k_a(\omega_0) - k_b(\omega_0') = 0 \) and the phase-matching function simplifies to:
\begin{eqnarray}
    \fl \qquad \qquad \Delta k(\nu, \nu') = \frac{d}{d \omega}k_p(\omega'_0 + \omega_0) (\nu' + \nu) - \frac{d}{d \omega}k_a(\omega_0) \, \nu - \frac{d}{d \omega} k_b(\omega_0') \, \nu'.
\end{eqnarray}
The three remaining parameters \(\frac{d}{d \omega}k_p(\omega'_0 + \omega_0)\), \(\frac{d}{d \omega}k_a(\omega_0)\) and \(\frac{d}{d \omega}k_b(\omega_0')\) --- the inverse group velocities of the three interacting beams --- define the material properties of the system and can be adjusted accordingly. 

This compact notation enables us to simulate any type-II PDC process with the help of just 6 parameters: The width and amplitude of the pump beam, the group velocities of the three interacting waves and the length of the nonlinear medium. 

In order to evaluate an (almost) uncorrelated type-II PDC case, with few optical modes \(r_k\), quite similar to the source discussed in \cite{eckstein_highly_2011}, as depicted in figure \ref{fig:pdc_time-ordering_result_uncorrelated}, we applied \(\sigma = 0.96231155\), \(\frac{d}{d \omega}k_p(\omega_0 + \omega_0') = 3.0 \), \(\frac{d}{d \omega}k_a(\omega_0) = 4.5 \), \(\frac{d}{d \omega}k_b(\omega_0') = 1.5\) and a crystal of length \(L = 2\). The pump amplitude \(E_p\) is adjusted to give the desired EPR squeezing values.

\section{PDC: Numerical implementation}\label{app:pdc_numerical_implementation}
In order to obtain the time-ordered solutions, we solved the classical differential equations in \eref{eq:pdc_classical_differential_equation_1} and \eref{eq:pdc_classical_differential_equation_2}, which give the functions \(U_a(z, \omega, \omega'')\), \(U_b(z, \omega, \omega'')\), \(V_a(z, \omega, \omega'')\), and \(V_b(z, \omega, \omega'')\) describing the type-II PDC process. 

In the numerical implementation of type-II PDC, we used a sampling of 500 points for each frequency degree of freedom and 500 points in \(z\)-direction to discretize the functions \(U_a(z, \omega, \omega'')\), \(U_b(z, \omega, \omega'')\), \(V_a(z, \omega, \omega'')\), \(V_b(z, \omega, \omega'')\) and \(f(\omega, \omega', z)\). As in the FC case, we evaluated the successive integrations via the trapezoid rule until the solutions converged. The actual solution defining the overall process properties is given by the matrices at the end of the crystal \(U_a(z = \frac{L}{2}, \omega, \omega'')\), \(U_b(z = \frac{L}{2}, \omega, \omega'')\), \(V_a(z = \frac{L}{2}, \omega, \omega'')\), and \(V_b(z = \frac{L}{2}, \omega, \omega'')\).

We checked the accuracy of the result in a variety of ways. At first we evaluated the canonical transformation conditions in \eref{eq:pdc_canonical_conditions_id}, \eref{eq:pdc_canonical_conditions_zero}, \eref{eq:pdc_inv_canonical_conditions_id}, and \eref{eq:pdc_inv_canonical_conditions_zero}. For example, in the case of \eref{eq:pdc_canonical_conditions_zero}, we evaluated
\begin{eqnarray}
    \nonumber
    \fl \qquad \int \mathrm d \omega' U_a(z = \frac{L}{2}, \omega, \omega') V_b(z = \frac{L}{2}, \omega'', \omega') \\
    \nonumber
    - \int \mathrm d \omega' V_a(z = \frac{L}{2}, \omega, \omega') U_b(z = \frac{L}{2}, \omega'', \omega') \\
    = D^{(diff)}(z = \frac{L}{2}, \omega, \omega')
\end{eqnarray}
and determined the distance of \(D^{(diff)}(z = \frac{L}{2}, \omega, \omega')\) from the expected zero matrix and consequently the error in the solution via
\begin{equation}
    \fl \qquad \mathrm{error} = \frac{\int \mathrm d \omega \int \mathrm d \omega' D^{(diff)}(z = \frac{L}{2},\omega, \omega')}{0.5 \left[ \int \mathrm d \omega \int \mathrm d \omega' |V_a(z = \frac{L}{2},\omega, \omega')| + \int \mathrm d \omega \int \mathrm d \omega' |U_b(z = \frac{L}{2},\omega, \omega')|\right]}.
\end{equation}
In all presented cases the obtained \(\mathrm{error}\) was below 0.000014.

We also checked the numerical Schmidt decompositions of \(U_a(z, \omega, \omega'')\), \(U_b(z, \omega, \omega'')\), \(V_a(z, \omega, \omega'')\), and \(V_b(z, \omega, \omega'')\) to verify the mode properties derived in \ref{app:PDC_canonical_transformation_conditions}. During this process the decompositions of \(U_a(z, \omega, \omega'')\), \(U_b(z, \omega, \omega'')\) showed numerical issues, these however could be resolved by decomposing \(U_a(z, \omega, \omega'') U_a^\dagger(z, \omega, \omega'')\), \(U_a^\dagger(z, \omega, \omega'') U_a(z, \omega, \omega'')\), \(U_b(z, \omega, \omega'') U_b^\dagger(z, \omega, \omega'')\), \(U_b^\dagger(z, \omega, \omega'') U_b(z, \omega, \omega'')\) instead. The obtained modes from these four matrices provided a much improved stability especially in the high-gain regime. Using these Schmidt modes we verified that the obtained Schmidt values of the \(U\) and \(V\) matrices behaved like \(\cosh(r_k)^2 - \sinh(r_k)^2 = 1\) with errors below 0.0002. We also asserted that the decompositions yielded the functions \(\varphi_k, \psi_k, \phi_k, \xi_k\) with symmetries as detailed in \eref{eq:pdc_svd_symmetries}, which where fulfilled within numerical accuracy.

The program code, written in Python, published on our website, is able to directly create the investigated type-II PDC processes and also performs all mentioned tests. It further enables the simulation of actual type-II down-conversion processes. In this case the unit for the length of the crystal has to match with the inverse group velocities and the unit for the width of the pump beam with the applied units for \(\nu\) and \(\nu'\).

\end{appendix}
\section*{References}
\bibliography{time-ordering}

\end{document}